\documentstyle[twocolumn,aps,prb,psfig]{revtex}

\begin{document}
  \draft

\title{Electron scattering states at solid surfaces calculated 
with realistic potentials}
\author{S.~Lorenz\cite{FHI}, C.~Solterbeck, and W.~Schattke}
\address{Institut f\"ur Theoretische Physik,
Christian-Albrechts-Universit\"at Kiel, Leibnizstra{\ss}e~15,
D-24098 Kiel, Germany}
\author{J.~Burmeister and W.~Hackbusch}
\address{ Mathematisches Seminar, Bereich 2,
Christian-Albrechts-Universit\"at Kiel, Hermann-Rodewald-Stra{\ss}e 3/1,
D-24098 Kiel, Germany}
\author{ }
\address{\copyright Copyright The American Physical Society 1997.  All rights
reserved.  Except as provided under U. S. copyright law, this work may not be
reproduced, resold, distributed or modified without the express permission of
The American Physical Society.  The archival version of this work was published
in Phys. Rev. B {\bf 55}, 13432 (1997).}
%\date{\today}

\maketitle
\thispagestyle{empty}
\begin{abstract}
Scattering states with low-energy-electron diffraction 
asymptotics are calculated for a general
non-muffin tin potential, as, e.g., for a pseudopotential with a 
suitable barrier and image potential part. The latter applies 
especially to the case of low lying conduction bands. The wave 
function is described with a reciprocal lattice
representation parallel to the surface and a discretization
of the real space perpendicular to the surface.
The Schr\"odinger equation leads 
to a system of linear one-dimensional equations.
The asymptotic boundary value problem is confined via the quantum
transmitting boundary method to a finite interval.
The solutions are obtained based on a multigrid technique
that yields a fast and reliable algorithm.
The influence of the boundary conditions, the accuracy, and the
rate of convergence with several solvers are discussed.
The resulting charge densities are
investigated.
\end{abstract}

%02.60.Lj Ordinary and partial differential equations; 
%          boundary value problems
%02.70.Bf Finite-difference methods
%61.14.Hg Low-energy electron diffraction (LEED) and reflection 
%          high-energy electron diffraction (RHEED)
%79.60.-i Photoemission and photoelectron spectra

Electron spectroscopies at low energies below $100$ eV are 
sensitive to the shape of the surface near potential. 
Sophisticated bandstructure calculations involve such potentials, 
but almost exclusively consider bound states. 
To obtain the same accuracy for current carrying states is still a 
challenge. Here we will consider final states of photoemission, 
which are time-reversed states of low energy electron diffraction 
(LEED). Typical energies of interest are $10$ eV 
and higher. With the advent of space-filling full-potential LEED 
calculations,\cite{leed2} the correct surface barrier can be 
included, but this requires, especially at these energies, 
extended CPU time. Another approach relies on smooth continuous 
matching\cite{matching} of the solutions inside the crystal to 
those in the vacuum. The inclusion of the surface potential by the 
propagation-matrix method~\cite{propagator} has proven to be an 
ill-posed problem.\cite{Wachutka85} 
The reason is the formulation as an initial value 
problem, which does not ensure the crucial continuous dependence on 
the boundary values. This is guaranteed by the two-side boundary 
conditions.\cite{Wachutka85,Wachutka:solution} Modern treatments of
elliptic problems often employ a discretization in direct space 
instead of choosing physically motivated basis functions.
A recent work on LEED and low energy positron diffraction 
presented a three-dimensional finite-difference method.\cite{joly} 
For the lower energies of ultraviolet photoemission spectroscopy,
the Coulomb singularities of the cores are less important. They can 
be avoided by using pseudopotentials, for which a mixed 
representation of the wave function seems to be suitable here.

In this contribution we address a direct, methodically simple, and fast 
solution of the Schr\"odinger equation with the scattering asymptotics 
treated as introduced by Lent and Kirkner.\cite{Lent90} The problem
is formulated as a two-side boundary problem.
The calculations refer exemplary to the GaAs (110) surface. 
The Schr\"odinger equation has to be solved for a given energy $E$ and 
a given direction of the electron incident on the surface with a
surface parallel wave vector $\vec{k}_{\|}$. The translation 
symmetry parallel to the surface allows a description of the wave 
function $\Psi$ and the potential $V$ in the Laue representation, 
consisting of a Fourier decomposition in the $xy$ plane parallel to 
the surface with the Fourier coefficients $\varphi_{\vec{g}}$ and 
$V_{\vec{g}}$ depending on the coordinate $z$ of the direct space
perpendicular to the surface,
\begin{eqnarray}
 \label{wavefunction}
 \Psi(\vec{\rho},z) &=& \sum_{\vec{g}} \varphi_{\vec{g}}(z) 
                        e^{i(\vec{k}_{\|}+
                        \vec{g} ) \cdot \vec{\rho}} \, ,\\
 \label{potentialgleichung}
 V(\vec{r}) &=& \sum_{\vec{g}} V_{\vec{g}}(z) e^{i\vec{g} \cdot 
                 \vec{\rho}}
              + \Sigma (E,z) \;.
\end{eqnarray}
The two-dimensional vector $\vec{\rho}$ lies in the $xy$ plane, the 
imaginary optical potential $\Sigma$ accounts for the attenuation 
owing to many-particle effects and other inelastic losses. The 
summation is over the reciprocal lattice vectors $\vec{g}$. 
In the Laue representation 
the Schr\"odinger equation appears as a system of linear 
one-dimensional differential equations, which we discretize by a 
grid with step size $h_z$
in the $z$ direction. This leads to the linear matrix equation
\begin{math}
 {\cal A}_{h_z} \vec{\varphi}_{h_z} = 0
\end{math}
with the solution $\vec{\varphi}_{h_z}$ and the coefficients
\begin{eqnarray}
 ({\cal A}_{h_z})_{i\vec g,i'\vec g'} &=&
  \left( \frac{- \delta_{i,i'+1} + 2
\delta_{i,i'}
   -\delta_{i,i'-1} } {h_{z}^2} \right) \delta_{\vec g,\vec g'} 
   \nonumber \\
    &+& \left[ \left( (\vec{g} + \vec{k_{\parallel}})^2
   +  \frac{2m}{\hbar^2}\left( -E + \Sigma (E,i)\right) \right)
      \delta_{\vec g,\vec g'} \right. \nonumber \\
    &+&  \left. \frac{2m}{\hbar^2} V_{\vec{g}'-\vec{g},i} \right] 
       \delta_{i,i'} \;.
\label{Matrixsystem}
\end{eqnarray}
The indices are the reciprocal lattice vectors $\vec{g}$, $\vec{g}'$ 
and the grid points $i$,$i'$. The complex matrix ${\cal A}_{h_z}$ is 
non-hermitean and indefinite. 
Since the potential relies on the Laue representation
without any restrictions, the method allows direct application of
potentials from modern band-structure computations. Nonlocal 
pseudopotentials can be applied in a way similar to their use in 
three-dimensional finite difference calculations of electronic 
structure.\cite{fdpm} For the calculation here we take as an 
example a potential, which was repeatedly applied to
photoemission calculations of GaAs(110).\cite{Henk93} It is a 
local pseudopotential with the surface
barrier taking into account relaxation, corrugation, 
and a smooth saturated image potential.

\begin{figure}\centering
\centerline{\psfig{figure=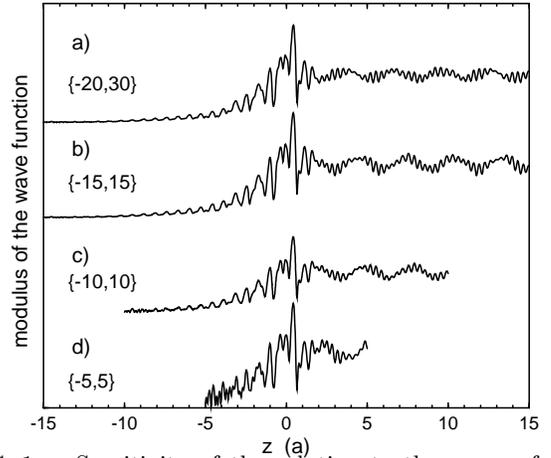,width=8.6cm,angle=270}}
   \caption{\label{z-Bereich} Sensitivity of the solution to the range 
    of calculation: wave functions at $(x,y) = (0,0)$ for different 
    grid sizes. The boundary positions are marked in $a$, at which 
    boundary conditions from a previous matching  calculation are used. 
    The potential for the matching was a step, here 
    it is the smooth barrier.} 
\end{figure}

Equation \ref{Matrixsystem} requires boundary conditions at 
the bulk and at the vacuum side of the grid.
The simplest model of a surface potential is a single step towards 
the vacuum, as used in 
matching calculations. A first guess for boundary conditions far away 
from the surface inside the crystal and that outside within the 
vacuum are Dirichlet values taken from such preliminary calculations.
In Fig. \ref{z-Bereich} the modulus of the wave functions for 
different grid sizes is plotted. 
Plots (c) and (d) show irregularities around the bulk boundary. Even 
with Neumann or mixed boundaries 
this feature cannot be suppressed. Because the damping reduces the 
wave functions to zero deep in the bulk, the required asymptotic 
behavior is fulfilled by taking  a zero value at a boundary 
sufficiently remote from the surface. Since the damping depends 
on energy, the coordinate of the boundary in the bulk should be 
chosen as energy dependent and automatically adjusted by tracking 
the neighboring values of the wave function.

In the vacuum the wave functions still differ even for high
grid sizes as shown in Fig. \ref{z-Bereich} (a) and (b). 
Therefore, we investigate this 
boundary more closely.
In a LEED experiment a beam of electrons is scattered at the surface 
of the crystal. The vacuum boundary condition has to guarantee
one single incoming plane wave. This can be tested by a half-sided
Fourier transformation of the wave function in the vacuum, 
assumin that the Fourier grid is well separated from the  
vicinity of the surface. 
The solution from a matching calculation does fulfill the
incoming beam asymptotics, the subsequent grid calculation with the 
same step potential does not.
The Fourier spectrum of the grid wave function 
shows additional incoming ghost waves that have contributions of 
up to $15\%$ and thereby strongly deteriorate the physical situation. 
Neumann and mixed boundaries from the matching do not improve this
unsatisfactory situation. Again the matching does not provide a 
correct boundary condition even far away from the crucial vicinity 
of the surface.

\begin{figure}\centering
\centerline{\psfig{figure=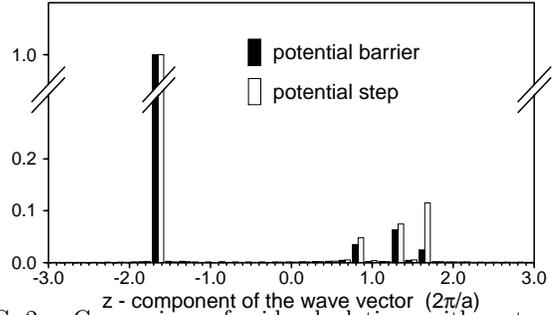,width=8.6cm,angle=270}}
 \caption{\label{fftcompare} Comparison of grid calculations with a
 step potential and a potential barrier: modulus of the 
 Fourier amplitudes normal to the surface in vacuum at $(x,y) = (0,0)$. 
 The squared amplitudes give the LEED-intensities.}
\end{figure}

\begin{figure}\centering
\centerline{\psfig{figure=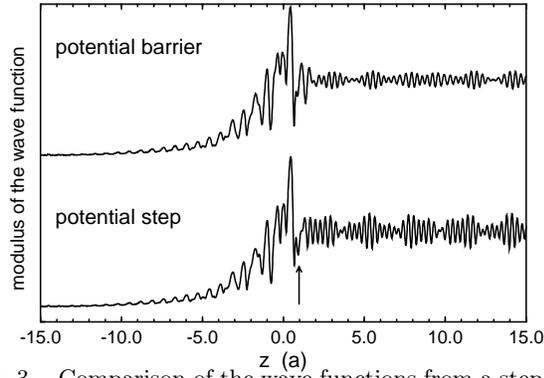,width=8.6cm,angle=270}}
 \caption{\label{wavecompare} Comparison of the wave functions from 
  a step potential and a potential barrier at $(x,y) = (0,0)$. The 
  position of the step is indicated by the arrow.}
\end{figure}

The correct boundary conditions in the vacuum are achieved by use 
of the elegant and simple quantum transmitting boundary method 
(QTBM).\cite{Lent90} The wave function in the vacuum is expanded 
into a set of propagating waves 
\begin{displaymath}   
 \Psi_{vac}{(\vec{\rho},z)} =  
 \sum_{\vec{g}} \left( \phi_{\vec{g}}^{(+)}  
 e^{+i{\kappa}_{\vec{g}}z}
 + \phi_{\vec{g}}^{(-)} 
 e^{-i{\kappa}_{\vec{g}}z} \right) 
 e^{i(\vec{k}_{\|}+\vec{g})\cdot\vec{\rho}} \; .
\end{displaymath}
$\phi_{\vec{g}}^{(+)}$ and $\phi_{\vec{g}}^{(-)}$ are the amplitudes
of the given incoming and unknown outgoing waves, respectively.
The grid wave function $\Psi_{grid}$ is that of 
Eq. (\ref{wavefunction}). For a short description of the quantum 
transmitting boundary method let $LN(\,)$ be a linear function and 
$FT(\,)$ the two-dimensional Fourier transformation. 
Continuity at boundary $z_r$ then yields at that plane
\begin{eqnarray*}
 \frac{\partial}{\partial z} \Psi_{vac} & = & LN \left( 
   \phi_{\vec{g}}^{(+)},
   \phi_{\vec{g}}^{(-)}  \right)
   =  LN \left( \phi_{\vec{g}}^{(+)}, FT(\Psi_{vac}) \right) \\
  & = &  LN \left( \phi_{\vec{g}}^{(+)}, FT(\Psi_{grid}) \right)
   =   LN \left( \phi_{\vec{g}}^{(+)}, \varphi_{\vec{g}}(z_r) 
     \right) \;.
\end{eqnarray*}
From the continuity of the normal derivative follows
\begin{displaymath}
   \frac{\partial}{\partial z} \Psi_{grid}  = 
   LN \left(\phi_{\vec{g}}^{(+)}, \varphi_{\vec{g}}(z_r) 
   \right)  \;  .
\end{displaymath}
This mixed linear boundary value problem is the QTBM that is 
inserted in Eq. (\ref{Matrixsystem}) with the derivative being 
discretized. The implementation of the QTBM is simple here due to 
the Laue representation. With the QTBM the calculated wave function 
fulfills the correct LEED asymptotics in the case of the step barrier 
as it does for any corrugated barrier too.
Figure \ref{fftcompare} shows the Fourier spectrum normal to the 
surface of a grid calculation with the QTBM basing either on a step 
potential or basing on the true barrier. 
The negative wave numbers correspond to incoming
waves. Both solutions show the postulated LEED asymptotics of one
incoming wave. The barrier potential influences strongly the solution, 
which is also illustrated by the
plot of the wave functions in Fig. \ref{wavecompare}.

Equation (\ref{Matrixsystem}) with the boundary conditions inserted 
gives a quadratic, linear, and inhomogeneous system of
equations, which has a block tridiagonal coefficient matrix. 
As an example we consider the GaAs(110) surface with a normal 
incident electron beam at an energy of $E_f=18$ eV, which in vacuum 
corresponds to a kinetic energy of $E_{kin}=12.75$ eV. The lateral 
plane wave basis of $57$ reciprocal lattice vectors proved to be 
sufficient for the employed potential. With the lattice constant of 
$a = 5.654$ \AA, the boundaries are chosen to be
at $\pm 15  a$, enclosing $45$ layers with $90$ atoms. The large 
distance from the surface to the vacuum boundary ensures a sufficient 
decay of the image potential. The positions of the boundaries deep 
inside the crystal and far outside in the vacuum cause a large grid, 
whose step size $h_z$ is bounded by several conditions. First, the 
wave function has to be correctly represented on the grid. This 
bound can be estimated by use of the sampling theorem. Furthermore, 
a small step size is needed for a stable discretization and a safe 
convergence of the iterative solvers. The Laplace operator has to 
dominate the zero-order terms, which require $h_z \le 0.06  a$ here.
Further bounds are given by the demands of applications. 
With $h_z= 0.012  a$ a typical 
number of $2500$ grid points and $142\,500$ equations results. 

Since the coefficient matrix results from an elliptic equation,  
is sparse, and is additionally a band matrix, a variety of 
solvers are available. As for the photoemission spectra a 
large series of final states has to be calculated, it is very 
important to reduce the CPU time per final state. Several direct and
iterative solvers have been tested. Some of the resulting CPU time
and memory requirements are given in Table \ref{cpu}. Time and memory 
increase linear with the number of grid points due to the simple 
structure of the coefficient matrix. The fastest direct solver was a 
routine in band storage mode performing an LU factorization that 
needs $98$ s for the test problem with a slope of 
$3.9\!\times\!10^{-2} s$ per grid point. It needs less than half of 
the time than a LEED calculation with space-filling potentials at 
this energy or a conventional matching at a potential step.

Iterative solvers in combination with multigrid methods are highly 
successful in solving differential equations.\cite{hackbusch} They have
recently been applied to the calculations of large-scale electronic 
structure.\cite{bernholc}
We implemented a two- and a three-grid method.
For the smoothing, the squared Jacobi iteration has been applied. With
a damping factor of 0.5 the best convergence rates have been obtained.
The Laue representation simplifies the implementation of the QTBM, 
but it causes strong codiagonals due to the potential 
coefficients. Therefore on the first level the equations must be 
treated by a direct solver, and the Jacobi iteration
is only used for smoothing on the finer grids. 
For the step size of $0.018  a$ on the coarse grid the two-grid
iteration
needs 20 steps to achieve the same mean defect of $\sim 10^{-14}$ 
as the direct solver. 
The three-grid methods do not improve the convergence rates.
The defect after 20~iterations was for 
a W cycle $6\!\times\!10^{-10}$ and for a V cycle $5\!\times\!10^{-7}$.
Theoretically the multigrid method needs $O(N)$ operations to solve a 
system of $N$ linear equations. The direct method for band matrices 
needs $O(Nw^2)$ operations. Because the number of codiagonals $w$ 
is the constant number of reciprocal lateral lattice vectors, both 
methods have the same asymptotic behavior. However, if a lower 
accuracy of the solution is acceptable, the multigrid method becomes
more favorable. If, e.g., a defect of $10^{-6}$ is sufficient, 
the two-grid method needs five iterations and $91$ s and gives an 
error in the modulus of the wave function of less than $0.2\%$.
Additionally, the maximum memory used can be reduced within the 
multigrid method. If the
system of equations is solved directly, a $(3w\!+\!1)\!\times\!N$   
matrix containing the LU factorization on the fine grid is needed. 
In the multigrid iteration with depth $l$ the $(2w\!+\!1)\!\times\!N$ 
matrix on the fine grid and the $(3w\!+\!1)\!\times\!(N 2^{-(l-1)})$ 
LU matrix on the coarse grid are stored. From the three-grid iteration 
onward, the required memory for the multigrid procedures becomes less 
than in the other methods. The multigrid calculations may possibly 
be accelerated by the use of iterative solvers better suited to the 
indefinite problem than the squared Jacobi iteration.

\begin{figure}\centering
\centerline{\psfig{figure=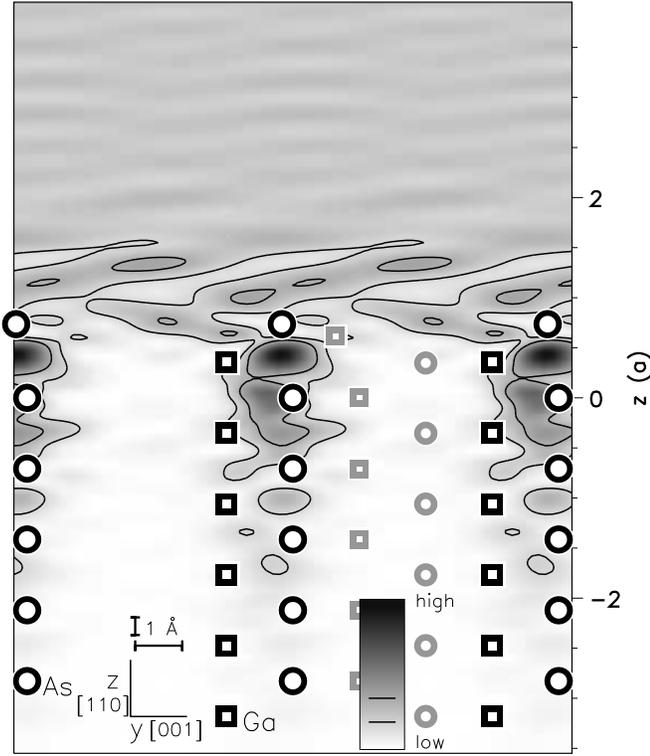,width=8.6cm}}
  \caption{\label{contour} Gray scale image and contour plot of the 
  charge density in the $yz$ plane at $x\!=\!0$ with a $z$ interval 
  from $-3.55 a$ to $3.95 a$. In the $y$ direction two unit cells 
  are shown. The atoms lying in the plane are drawn black, and in the 
  right cell the projected positions of the remaining atoms are gray. 
  The different scales for the $y$ and $z$ directions are indicated 
  by the {\AA}ngstr{\o}m scale.
  }
\end{figure}

As a first application the charge-density distribution of a LEED state 
is shown in Fig. \ref{contour}. 
The interface crystal vacuum and the exponential 
decay of the wave function into the crystal are clearly visible. 
In the vacuum a distance of several lattice constants 
from the surface is necessary to evolve 
an interference pattern typical of superimposed waves, c.f.\  
Fig. \ref{wavecompare}. 
The importance of the correct treatment of the potential barrier, which
is already obvious from the different LEED intensities 
in Fig. \ref{fftcompare}, is illustrated by the strong 
charge fluctuations in the surface region. In contrast to tracking 
multiple scattering paths in conventional LEED calculations, the 
properties of LEED states are much easier interpreted with a single 
wave function. It is interesting to see that there are well localized 
regions of high charge density even at this nonbonding energy. In 
applications to photoemission these regions will give strong 
contributions to the photocurrent. Thus, this treatment supports a 
local interpretation of the emission. First photoemission calculations 
with these final states were promising.

Solutions of the Schr\"odinger equation with scattering boundary 
conditions have been calculated with a realistic potential of the 
GaAs (110) surface. The correct asymptotic behavior was obtained 
by the quantum transmitting boundary method. The Laue representation 
allows a simple implementation of the boundary conditions. Several 
multigrid methods and direct solvers have been tested. The fastest 
were a routine in band storage mode with LU factorization 
and, when stopping at a slightly lower accuracy, a two-grid 
method. The multigrid method is competitive with direct solvers due 
to its higher flexibility. A general implementation of the potential 
allows us to use potentials from modern {\em ab-initio} techniques. In an 
application, the calculated charge density of a LEED state showed 
distributions with localized accumulation regions of the excited 
electron. It may yield a comfortable access to a local interpretation 
of photoemission spectra.

This work was supported by the  Bundesministerium 
f\"{u}r Bildung, Wissenschaft, Forschung und Technologie.

\begin{table}
 \Large
 \begin{center}
  \begin{tabular}{lcc}
   \hline
 Method           &  CPU time (s) &  Memory (MB)  \\ \hline \hline
 Sparse Gauss               &  1862          & 747  \\
 Sparse LU                  & (1862)     & $>$2000  \\
 Band LU with refinement    &   257          & 654  \\ 
 Band LU without refinement &    98          & 392  \\ 
 Two grid                   &   208          & 458  \\ 
 Three grid V cycle         &   343          & 360  \\
 Three grid W cycle         &   270          & 360  \\ 
\end{tabular}
\end{center} 
\caption{\label{cpu} CPU time and memory used by different direct
       and iterative routines for a system of $142\,500$~equations on a
       CRAY-J916. As direct solvers routines from the International Mathematics
       and Scientific Library were
       used. For the routine with the LU factorization of a sparse 
       matrix the  time is extrapolated from tests with smaller grids. 
       The given memory is required by the $64$-bit architecture of the 
       Cray.
}
\end{table}

\end{document}